\documentclass[]{article}

\usepackage{fullpage}
\usepackage{authblk}
\usepackage{amsfonts}
\usepackage{multirow}
\usepackage{mathrsfs}
\usepackage{graphicx}
\usepackage{amsmath}
\usepackage{amssymb}
\usepackage{bm}
\usepackage{bbm}
\usepackage{color}
\usepackage{slashed}
\usepackage{diagbox}
\usepackage{ulem}
\usepackage{wrapfig}
\usepackage[hidelinks]{hyperref}
\usepackage{cleveref}
\usepackage{fancyhdr}
\usepackage{subcaption}
\numberwithin{equation}{section}

\begin{document}

\fancypagestyle{plain}{
    \fancyhead[R]{}
    \renewcommand{\headrulewidth}{0pt}
}

\title{\bf{\textsf{Reduction of Feynman Integrals in the Parametric Representation III: Integrals with Cuts}}}

\author[a]{\bf{\textsf{Wen Chen}}\footnote{wchen1@ualberta.ca}}
\affil[a]{\small{\it{Department of Physics, University of Alberta, Edmonton, AB, T6G 2E1, Canada}}}

\date{\small{\today}}

\maketitle

\begin{abstract}

Phase space cuts are implemented by inserting Heaviside theta functions in the integrands of momentum-space Feynman integrals. By directly parametrizing theta functions and constructing integration-by-parts (IBP) identities in the parametric representation, we provide a systematic method to reduce integrals with cuts. Since the IBP method is available, it becomes possible to evaluate integrals with cuts by constructing and solving differential equations.

\end{abstract}


\section{Introduction}

Feynman integrals with cuts are frequently encountered in perturbative calculations in high energy physics, especially while calculating various jet observables and event-shape distributions. Generally, cuts are implemented by inserting Heaviside theta functions in the integrands in the momentum space. The presence of theta functions largely complicates the calculations of Feynman integrals.

The most widely used technique to reduce Feynman integrals is the integration-by-parts (IBP) method \cite{Tkachov:1981wb,Chetyrkin:1981qh}. However, it is not clear how to directly apply the regular IBP method to integrals with cuts. In a recent paper \cite{Baranowski:2020xlp}, theta functions were written as integrals of delta functions. The resulting integrals were reduced by combining the reverse unitarity \cite{Anastasiou:2002yz} and the IBP method. By using this method, one has to introduce an extra scale for each theta function. Consequently, the reduction becomes much more complicated for integrals with several cuts. Thus the application of this method to more complicated integrals is far from trivial.

On the other hand, it was suggested that IBP identities can directly be derived in the parametric representation \cite{Lee:2013hzt,Lee:2014tja}. It can be shown that each momentum-space IBP identity \cite{Baikov:1996iu} corresponds to a shift relation in the parametric representation \cite{Bitoun:2017nre}. Since a theta function has an integral representation quite similar to the Schwinger parametrization of a propagator, it is possible to directly parametrize theta functions and construct IBP identities in the parametric representation. In this paper, we show that the methods developed in Refs. \cite{Chen:2019mqc,Chen:2019fzm} (referred to as paper I and paper II, respectively, hereafter) to parametrize and reduce tensor integrals can be applied to integrals with theta functions with slight modifications.

This paper is organized as follows. In \cref{Sec:ParIBP}, we show how to use the method developed in paper I and paper II to parametrize integrals with cuts and to construct IBP identities for them. Some detailed examples are provided in \cref{Sec:Exampl}.

\section{Parametrization and IBP identities}\label{Sec:ParIBP}

It is well-known that a propagator can be parametrized by

\begin{equation}
\frac{1}{D_i^{\lambda_i+1}}=\frac{e^{-\frac{\lambda_i+1}{2}i\pi}}{\Gamma(\lambda_i+1)}\int_0^{\infty}\mathrm{d}x_i~e^{ix_iD_i}x_i^{\lambda_i},\qquad\text{Im}\{D_i\}>0.
\end{equation}

\noindent Heaviside theta functions have a similar integral representation

\begin{equation*}
\theta(D_i)=-\frac{i}{2\pi}\int_{-\infty}^{\infty}\mathrm{d}x_i\frac{e^{ixD_i}}{x_i+i0^+}.
\end{equation*}

\noindent For future convenience, we define the function

\begin{equation}\label{Eq:ThetPar}
w_\lambda(u)\equiv e^{-\frac{\lambda+1}{2}i\pi}\int_{-\infty}^{\infty}\mathrm{d}x\frac{1}{x^{\lambda+1}}e^{ixu}.
\end{equation}

\noindent It's easy to see that

\begin{align*}
w_0(u)=&2\pi\theta(u),\\
w_{-1}(u)=&2\pi\delta(u),\\
w_{-2}(u)=&2\pi\delta^\prime(u).
\end{align*}

\noindent With this representation, the standard procedure to parametrize Feynman integrals can easily be generalized to integrals with theta functions. Following the convention used in paper I and paper II, we have

\begin{equation}
\begin{split}
M\equiv&\pi^{-\frac{1}{2}Ld}\int d^dl_1d^dl_2\cdots d^dl_L\frac{w_{\lambda_1}(D_1)w_{\lambda_2}(D_2)\cdots w_{\lambda_m}(D_m)}{D_{m+1}^{\lambda_{m+1}+1}D_{m+2}^{\lambda_{m+2}+1}\cdots D_n^{\lambda_n+1}}\\
=&s_g^{-\frac{L}{2}}e^{i\pi\left[\lambda_{n+1}-\frac{d}{2}+1-\sum_{i=1}^m(\lambda_{i}+\frac{1}{2})\right]}I(\lambda_0,\lambda_1,\ldots,\lambda_n),
\end{split}
\end{equation}

\noindent where $s_g$ is the determinant of the $d$-dimensional metric, and $\lambda_{n+1}\equiv-(L+1)\lambda_0-1+\sum_{i=1}^m\lambda_i-\sum_{i=m+1}^n(\lambda_i+1)$, with $\lambda_0\equiv-\frac{d}{2}$. We have the parametric integral

\begin{equation}\label{Eq:ParInt}
\begin{split}
I(\lambda_0,\lambda_1,\ldots,\lambda_n)\equiv&\int \mathrm{d}\Pi^{(n+1)}\mathcal{I}^{(-n-1)}\\
\equiv&\frac{\Gamma(-\lambda_0)}{\prod_{i=m+1}^{n+1}\Gamma(\lambda_i+1)}\int \mathrm{d}\Pi^{(n+1)}\mathcal{F}^{\lambda_0}\prod_{i=1}^{n+1}x_i^{\lambda_i}.
\end{split}
\end{equation}

\noindent Here the measure is $\mathrm{d}\Pi^{(n)}\equiv \prod_{i=1}^{n+1}\mathrm{d}x_i\delta(1-\sum_j|x_j|)$, where the sum in the delta function runs over any nontrivial subset of $\{x_1,x_2,\ldots,x_{n+1}\}$. The polynomial $\mathcal{F}(x)\equiv F(x)+U(x)x_{n+1}$. $U$ and $F$ are Symanzik polynomials, defined by $U(x)\equiv\det{A}$, and $F(x)\equiv U(x)\left(\sum_{i,j=1}^L(A^{-1})_{ij}B_i\cdot B_j-C\right)$. Polynomials $A$, $B$, and $C$ are defined through $\sum_{i=1}^nx_iD_i\equiv\sum_{i,j=1}^LA_{ij}l_i\cdot l_j+2\sum_{i=1}^LB_i\cdot l_i+C$.

It should be noticed that in the definition of the parametric integral in \cref{Eq:ParInt}, for a ``propagator" $w_{\lambda_i}(D_i)$, there is no corresponding gamma function in the prefactor. And the corresponding index $\lambda_i$ can be both positive and negative.

Similar to the parametric IBP identities derived in paper I, we have

\begin{subequations}
\begin{align}
0=&\int \mathrm{d}\Pi^{(n+1)}\frac{\partial}{\partial x_i}\mathcal{I}^{(-n)},&& i=1, 2,\ldots, m,\\
0=&\int \mathrm{d}\Pi^{(n+1)}\frac{\partial}{\partial x_i}\mathcal{I}^{(-n)}+\delta_{\lambda_i0}\int \mathrm{d}\Pi^{(n)}\left.\mathcal{I}^{(-n)}\right|_{x_i=0},&& i=m+1, m+2,\ldots, n+1.
\end{align}
\end{subequations}

We define the index-shifting operators $R_i$, $D_i$, and $A_i$, with $i=0,1,\dots,n$, such that

\begin{align*}
R_iI(\lambda_0,\dots,\lambda_i,\dots,\lambda_n)=&(\lambda_i+1)I(\lambda_0,\dots,\lambda_i+1,\dots,\lambda_n),\\
D_iI(\lambda_0,\dots,\lambda_i,\dots,\lambda_n)=&I(\lambda_0,\dots,\lambda_i-1,\dots,\lambda_n),\\
A_iI(\lambda_0,\dots,\lambda_i,\dots,\lambda_n)=&\lambda_iI(\lambda_0,\dots,\lambda_i,\dots,\lambda_n).
\end{align*}

\noindent It is understood that

\begin{equation*}
I(\lambda_0,\dots,\lambda_{i-1},-1,\dots,\lambda_n)\equiv\int \mathrm{d}\Pi^{(n)}\left.\mathcal{I}^{(-n)}\right|_{x_i=0},\quad i=m+1,~m+2,~\cdots,~n.
\end{equation*}

\noindent We formally define operators $D_{n+1}$ and $R_{n+1}$, such that $D_{n+1}I=I$, and $R_{n+1}^iI=(A_{n+1}+1)(A_{n+1}+2)\cdots(A_{n+1}+i)I$, with $A_{n+1}\equiv-(L+1)A_0+\sum_{i=1}^mA_i-\sum_{i=m+1}^n(A_i+1)$. We further introduce the operators $\hat{x}_i$, $\hat{z}_i$ and $\hat{a}_i$ such that

\begin{align*}
\hat{x}_i=&\left\{
\begin{matrix}
D_i&,&i=1,~2,\ldots,~m,\\
R_i&,&i=m+1,~m+2,\ldots,~n+1,
\end{matrix}
\right.\\
\hat{z}_i=&\left\{
\begin{matrix}
-R_i&,&i=1,~2,\ldots,~m,\\
D_i&,&i=m+1,~m+2,\ldots,~n+1,
\end{matrix}
\right.\\
\hat{a}_i=&\left\{
\begin{matrix}
-A_i-1&,&i=1,~2,\ldots,~m,\\
A_i&,&i=m+1,~m+2,\ldots,~n+1.
\end{matrix}
\right.
\end{align*}

\noindent Obviously we have $\hat{a}_{n+1}=-(L+1)A_0-\sum_{i=1}^n(\hat{a}_i+1)$. For $i=1,~2,\ldots,n$, we have the following commutation relations:

\begin{align*}
\hat{z}_i\hat{x}_j-\hat{x}_j\hat{z}_i=&\delta_{ij},\\
\hat{z}_i\hat{a}_j-\hat{a}_j\hat{z}_i=&\delta_{ij}\hat{z}_i,\\
\hat{x}_i\hat{a}_j-\hat{a}_j\hat{x}_i=&-\delta_{ij}\hat{x}_i.
\end{align*}

\noindent With the operators $\hat{x}_i$, $\hat{z}_i$, and $\hat{a}_i$, it is easy to write the IBP identity in the following form

\begin{equation}
D_0\frac{\partial\mathcal{F}(\hat{x})}{\partial \hat{x}_i}-\hat{z}_i\approx 0,\quad i=1,~2,\dots,n+1.
\end{equation}

\noindent Here we use $\approx$ to emphasize that these equations are valid only when they are applied to nontrivial parametric integrals.

The methods developed in paper II to parametrize tensor integrals and to construct dimensional-shift-free parametric IBP identities can easily be applied to integrals with cuts. One only needs to do the replacements $R_i\to\hat{x}_i$, $D_i\to\hat{z}_i$, and $A_i\to\hat{a}_i$. Differential equations can also be constructed by using eq. (3.18) in paper II. Here we do not need to go into detail. Thus, in principle, integrals with cuts can be evaluated by using the standard differential-equation method \cite{Kotikov:1990kg,Remiddi:1997ny,Gehrmann:1999as,Henn:2013pwa,Lee:2014ioa}.

\section{Examples}\label{Sec:Exampl}
 
 \begin{figure}[ht]
 \centering
\begin{subfigure}{.25\textwidth}
  \centering
  \includegraphics[width=.8\linewidth]{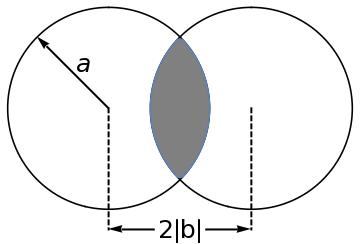}  
  \caption{}
  \label{Fig:figa}
\end{subfigure}
\begin{subfigure}{.25\textwidth}
  \centering
  \includegraphics[width=.8\linewidth]{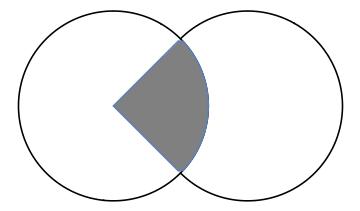}  
  \caption{}
  \label{Fig:figb}
\end{subfigure}
\begin{subfigure}{.25\textwidth}
  \centering
  \includegraphics[width=.8\linewidth]{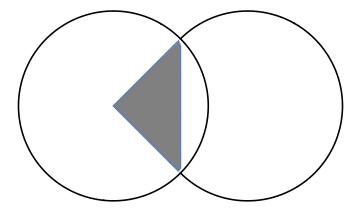}  
  \caption{}
  \label{Fig:figc}
\end{subfigure}
\caption{The geometric interpretation of \cref{Eq:RedInt1}.}
\label{Fig:fig}
\end{figure}

We first consider the following simple but interesting example.

\begin{equation*}
I_1(-\frac{d}{2},\lambda_1,\lambda_2)\equiv \frac{i}{\pi^{d/2}}\int d^dr~w_{\lambda_1}(a^2-r^2)~w_{\lambda_2}(a^2-(r-2b)^2)
\end{equation*}

\noindent By using the method I described in paper II (cf. eq. (3.17) therein), we get the following IBP identities.

\begin{align*}
A_1-A_2-4 b^2 D_1+4 b^2 D_2+D_2 R_1-D_1 R_2\approx&0,\\
2 A_0-2 A_1-A_2+2 a^2 D_1+2 a^2 D_2-4 b^2 D_2-D_2 R_1\approx&0.
\end{align*}

Specifically, we consider the reduction of the integral $I_1(-d/2,0,0)$. By solving IBP identities, we get

\begin{equation}\label{Eq:RedInt1}
\begin{split}
I_{1a}\equiv&-\frac{i}{4}\pi^{\frac{d}{2}-2}I_1(-d/2,0,0)\\
=&\int d^dr~\theta(a^2-r^2)~\theta(a^2-(r-2b)^2)\\
=&\frac{4a^2}{d}\int d^dr~\delta(a^2-r^2)~\theta(a^2-(r-2b)^2)\\
&-\frac{16b^2(a^2-b^2)}{d(d-1)}\int d^dr~\delta(a^2-r^2)~\delta(a^2-(r-2b)^2)\\
\equiv&\frac{4a^2}{d}I_{1b}-\frac{16b^2(a^2-b^2)}{d(d-1)}I_{1c}.
\end{split}
\end{equation}

\noindent This result has an interesting geometric interpretation. It is easy to see that the integral $I_{1a}$ is nothing but the volume of the intersection of two $d$-dimensional balls with a radius $a$ separated by a distance of $2|b|$, as is shown in \cref{Fig:figa}. $2aI_{1b}=\int d^dr~\delta(a-r)~\theta(a^2-(r-2b)^2)$ is the bottom area of the $d$-dimensional cone shown in \cref{Fig:figb}. Thus $\frac{2a^2}{d}I_{1b}$ is the volume of this $d$-dimensional cone. Similarly, $8b\sqrt{a^2-b^2}I_{2c}$ is the perimeter of the intersection of two spheres (the surfaces of the two balls). This will become obvious by using azimuthal coordinates. Thus $\frac{8b^2(a^2-b^2)}{d(d-1)}I_{1c}$ is the volume of the $d$-dimensional cone (with a flat bottom) shown in \cref{Fig:figc}. Hence \cref{Eq:RedInt1} just tells us how to calculate the volume of the intersection of two balls.

We can also construct differential equations for these integrals. The differential operator reads (cf. eq. (3.18) in paper II)

\begin{equation*}
\frac{\partial}{\partial b^2}=\frac{1}{2b^2}A_2-\frac{1}{2b^2}D_2R_1-2D_2.
\end{equation*}

\noindent Applying this operator to the integrals $I_{1b}$ and $I_{1c}$, and carrying out IBP reductions, we get the following differentiation equations:

\begin{equation*}
\frac{\partial}{\partial b^2}
\begin{pmatrix}
I_{1b}\\
I_{1c}
\end{pmatrix}
=
\begin{pmatrix}
0&-2\\
0&-\frac{a^2+(d-4)b^2}{2b^2(a^2-b^2)}
\end{pmatrix}
\begin{pmatrix}
I_{1b}\\
I_{1c}
\end{pmatrix}.
\end{equation*}

\noindent It is easy to check that the solutions of these equations do agree with the result obtained by a direct calculation.

As a less trivial example, we consider the reduction of the integral

\begin{equation*}
I_2=\frac{(2\pi)^6}{\pi^d}\int d^dl_1 d^dl_2\frac{\delta(l_1^2)\delta(l_2^2)\delta(l_1^+-a)\delta(l_2^--b)\theta(l_1^--l_1^+)\theta(l_2^+-l_2^-)}{l_1^+l_1^-(l_1^++l_2^+)(l_1^-+l_2^-)}.
\end{equation*}

\noindent Here the lightcone coordinates are used. That is, $l_i^+\equiv l_i\cdot n$, and $l_i^-\equiv l_i\cdot\bar{n}$, with $n^2=\bar{n}^2=0$, and $n\cdot\bar{n}=2$. This integral is relevant for the calculation of the two-loop hemisphere soft functions \cite{Kelley:2011ng}. This integral can be reduced to

\begin{align*}
I_2=&-\frac{2}{(d-4)ab}\frac{(2\pi)^6}{\pi^d}\int d^dl_1 d^dl_2\frac{\delta(l_1^2)\delta(l_2^2)\delta(l_1^+-a)\delta(l_2^--b)\delta(l_1^--l_1^+)\theta(l_2^+-l_2^-)}{l_1^++l_2^+}\\
&-\frac{1}{ab}\frac{(2\pi)^6}{\pi^d}\int d^dl_1 d^dl_2\frac{\delta(l_1^2)\delta(l_2^2)\delta(l_1^+-a)\delta(l_2^--b)\theta(l_1^--l_1^+)\theta(l_2^+-l_2^-)}{(l_1^++l_2^+)(l_1^-+l_2^-)}.
\end{align*}

\noindent The detailed calculation is carried out by using a home-made Mathematica code. We have verified this result by explicit calculations of these integrals.

To validate our method, we have also applied this method to some practical calculations. For example, we reproduce the decay rate for the four-lepton decay $\gamma^*\to l\bar{l}l\bar{l}$, which can be obtained from the decay rate of the four-quark decay $\gamma^*\to q\bar{q}q\bar{q}$ \cite{Gehrmann-DeRidder:2003pne,GehrmannDeRidder:2004tv} by stripping off some color factors. The detailed calculation is carried out as follows. We first generate IBP identities by using the method described in this paper. Then we solve these identities by using the package \texttt{Kira} \cite{Klappert:2020nbg}.

\section{Summary}

By directly parametrizing Heaviside theta functions and constructing IBP identities in the parametric representation, we provide a systematic method to reduce integrals with cuts. We show that the methods developed in paper I and paper II to parametrize and to reduce regular Feynman integrals can be applied to integrals with cuts by slightly modifying the definitions of the index-shifting operators. Differential equations can also be constructed. Thus, in principle, the standard differential equation method can be used to evaluate integrals with cuts.

\section*{\normalsize{Acknowledgments}}

The author thanks for the hospitality of the Institute of High Energy Physics, Chinese Academy of Sciences, where part of this work was finished. This work was supported by the Natural Sciences and Engineering Research Council of Canada.

\end{document}